\documentclass[linenumbers]{aastex631}

\newcommand{\Msun}{$\textrm{M}_\odot$}
\newcommand{\Rstar}{$\textrm{R}_\star\,$}

\begin{document}
\nolinenumbers
\title{Catalog of magnetic white dwarfs with hydrogen dominated atmospheres}

\author{L. L. Amorim}
\affiliation{Instituto de Física, Universidade Federal do Rio Grande do Sul, 91501-900 Porto-Alegre, RS, Brazil}

\author{S. O. Kepler}
\affiliation{Instituto de Física, Universidade Federal do Rio Grande do Sul, 91501-900 Porto-Alegre, RS, Brazil}


\author{Baybars Külebi}
\affiliation{Institut de Ciencies de L’Espai, Universitat Autonoma de Barcelona and Institute for Space Studies of Catalonia, Gran Capita 2-4,
Edif. Nexus 104, E-08034 Barcelona, Spain}

\author{S. Jordan}
\affiliation{Astronomisches Rechen-Institut, Zentrum fur Astronomie der Universität Heidelberg, Mönchhofstr. 12-14, D-69120 Heidelberg, Germany}

\author{A. D. Romero}
\affiliation{Instituto de Física, Universidade Federal do Rio Grande do Sul, 91501-900 Porto-Alegre, RS, Brazil}





\begin{abstract}
White dwarfs are excellent research laboratories as they reach temperatures, pressures, and magnetic fields that are unattainable on Earth. To better understand how these three physical parameters interact with each other and with other stellar features, we determined the magnetic field strength for a total of 804 hydrogen-rich white dwarfs of which 287 are not in the literature. We fitted the spectra observed with the Sloan Digital Sky Survey using atmospheric models that consider the Zeeman effect due to the magnetic field at each point in the stellar disk. Comparing magnetic and non-magnetic WDs, the literature already shows that the magnetic ones have on average higher mass than the non-magnetic. In addition to that, magnetic fields are more common in cooler WDs than in hotter WDs. In consonance, we found that those with higher magnetic field strengths tend to have higher masses, and lower temperatures, for which models indicate the crystallization process has already started. This reinforces the hypothesis that the field is being generated and/or amplified in the cooling process of the white dwarf. Our sample constitutes the largest number of white dwarfs with determined magnetic fields to date.

\end{abstract}



\section{Introduction} \label{sec:intro}

Strong magnetic fields are common in Hydrogen-rich white dwarfs stars (DAs). \cite{2013Kepler} showed that at least 4\% of all DAs observed with \textit{Sloan Digital Sky Survey} (SDSS) until Data Release 7 have magnetic fields greater than 1 MG. The authors visually inspected all DA spectra and found 521 stars with Zeeman Splittings. A more robust method to determine the field was presented by \cite{2009Kulebi}, who used least squares minimization to find the best model of magnetic field geometry to fit the observed spectra. They applied this technique to 141 magnetic white dwarfs rich in hydrogen (DAH). A revisit to the previous measurements is insightful, since the largest work with DAHs considered only the splitting in H-alpha and H-beta and used only a visual field determination method. We want complete and homogeneous results to try to understand general properties of these stars to add information to the question of the origin of the magnetic fields. In this work, we use the fitting method to measure the magnetic field of all 804 DAHs found in the SDSS sample until Data Release 16 (DR16). From this, 287 are newly reported DAHs. 

The fraction of detected magnetic white dwarfs depends on the observed sample, the detection method, and spectral types. \cite{2021Bagnulo} found 23.4 ± 4.8 \% of DAHs in the DA volume-limited sample for 20 pc using spectropolarimetry, while \cite{2013Kepler} encountered 4\% of DAHs in the DAs of a SDSS magnitude limited sample with spectroscopy only.  \cite{2020Kawka} presents a table of differences in the fraction of magnetic white dwarfs through various spectral types. The fraction of detected magnetic white dwarfs may differ from the real fraction of magnetic white dwarfs due to limitations on the domain of magnetic field strength studied, the quality of the data, or the significance of the measurable physical effect.

The origin of magnetic fields in white dwarfs is still an open question after more than fifty years of the first discovery by \cite{1970Kemp}. A systematic study is crucial for the construction of a significant statistical sample, which may help us shed some light on their origin. The magnetic field could be formed in three different stages of the white dwarf evolution: before the white dwarf stage, during its formation, or during the cooling process.

The main hypothesis of the first group corresponds to the fossil fields from Ap/Bp stars (chemically peculiar and with magnetic fields stronger than classical A- or B-type stars). The original gas from which stars are formed are probably magnetized, since the net magnetic field of the galaxy is not zero. In the main sequence, these fields are usually small, of the order of a few kG as first shown by \cite{1947Babcock} and later examples in \cite{1958Babcock}. The magnetic field can be boasted through conservation of the magnetic flux up to 100~MG when the star gets stripped of its outer layers and its core gets exposed and starts to contract during the white dwarf cooling sequence.

This possible origin of the magnetic field is very attractive because the Ohmic decay in degenerate matter suggests that these fields should last billions of years. However, \cite{2005WickFerrario} concluded that the amount of magnetic Ap/Bp stars that have been detected cannot account for the fraction of magnetic white dwarfs (MWD) measured, so other mechanisms must also occur.

For the magnetic field to arise during the formation of the white dwarf, the system may not be single. It can be due to the merger of two degenerate cores, or it can be formed during the interaction of the two components of the binary (common envelope) as presented by \citet[e.g.][]{2008Tout}. However, this channel of magnetism formation would lead to a much higher magnetic incidence among white dwarfs in close binaries than is currently observed \citep[e.g][]{2020Belloni}. 

\cite{1988Liebert} was the first to suggest that magnetic WDs have generally a higher mass than non-magnetic WDs. This was consistently found in several subsequent works, such as \cite{2013Kepler} and \cite{2020McCleery}. \cite{2021Bagnulo} argued that the fact that the magnetic have, in average, a higher mass than the non-magnetic is true only for young WDs. The possible origins mentioned so far are in agreement with the higher mass that MWDs have when compared to the whole sample. 

Nevertheless, there is evidence of yet another way of forming magnetic fields in white dwarfs. It was early supposed that magnetic fields are more common and stronger in cooler white dwarfs because it was first detected in this group (\cite{1971Greenstein}). Eventually, hot magnetic white dwarfs were also detected, and the early result was indeed corroborated (\cite{1979Liebert}). The sample was still subjected to strong selection effects and the results were questioned. \cite{1999Valyavin} studied the evolution of magnetic white dwarfs in an even larger sample and concluded that as the star cools, the frequency of magnetic white dwarfs increases, as does the strength of the magnetic fields detected. This was endorsed by further studies such as \cite{2013Kepler} and \cite{2021Bagnulo}. This means that the white dwarf must be producing, exposing or enhancing the surface magnetic field.

When a DA white dwarf cools below 14\,000 K, it develops a surface convective layer in which the dynamo process can occur, giving a boost to the surface field. However, as the temperature continues to drop, the kinetic energy of the envelope becomes larger, and eventually the convective cells hinder the magnetic field line movement. This is coherent with the further drop of magnetic field strength at even lower temperatures.

We could also mention other possibilities that could account for the magnetic field in white dwarfs, like the crystallization of its core \citep{2017Isern,2022Ginzburg} and the interaction with orbiting planets \citep{2021Schreiber}.The later effect will not be further studied in this work, as we have yet no evidence that it is statistically significant to the complete sample.

\section{Detection of magnetic fields in SDSS DR16 white dwarfs}

To build our sample, we visually investigated all DA spectra from Sloan Digital Sky Survey (SDSS) DR7 to DR16 searching for Zeeman Splittings and concatenated the selected ones with the previously known until Data Release 7, resulting in a total of 804 magnetic white dwarfs with SDSS spectra. We used the code (YAWP) presented by \cite{2009Kulebi} to determine the strength of the magnetic fields across the surface of the star which better matches the observed spectrum, assuming an off-center dipole inclined in relation to our line of sight. 

\begin{figure}[h]
    \centering
	\includegraphics[width=0.85\linewidth]{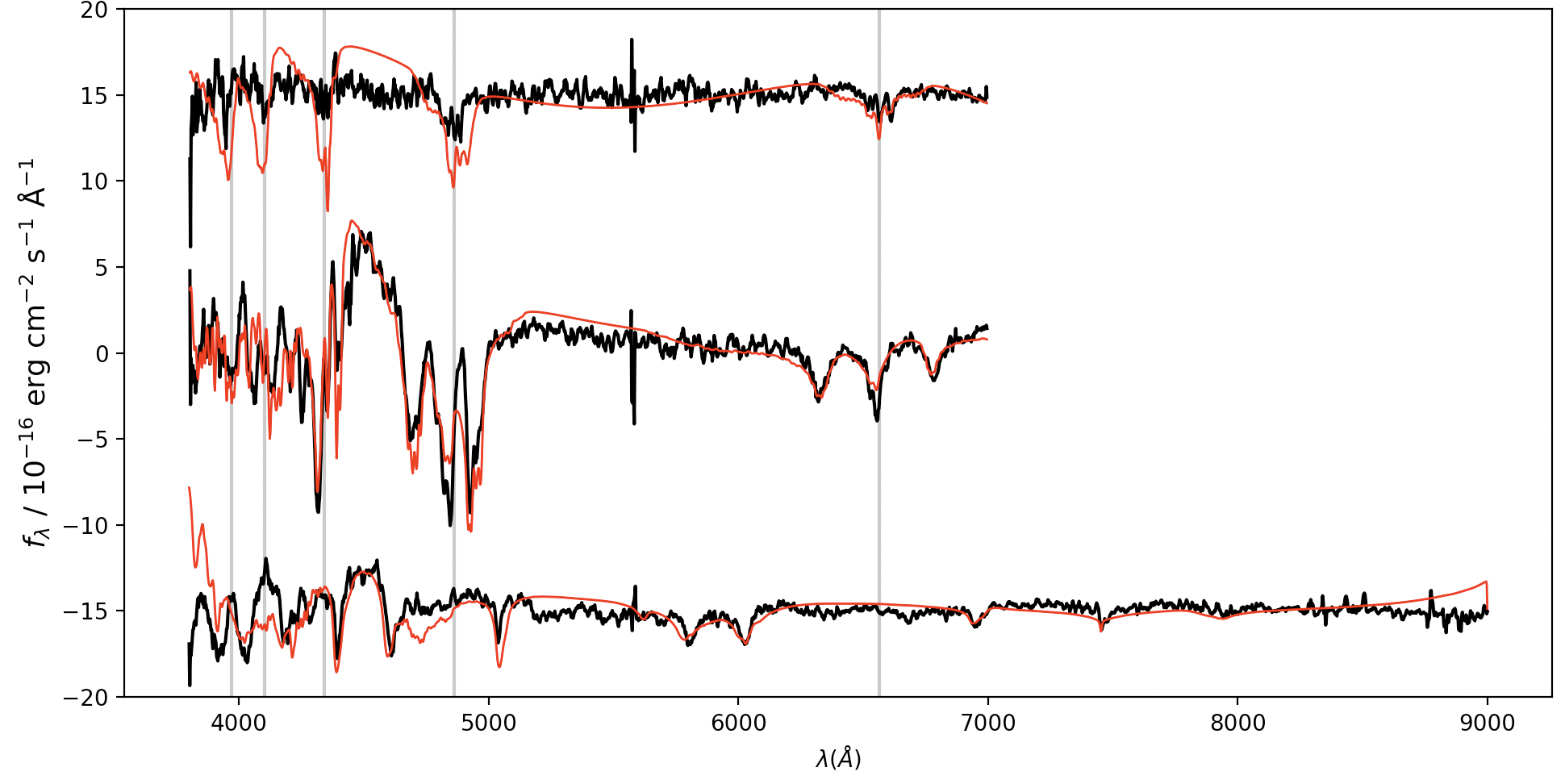}
    \caption{SDSS spectrum of three of the stars we computed the magnetic field. The model with the best least-squares fit to the observed data is shown in red. From top to bottom we have (a) SDSS J101529.62+090703.7, Plate-MJD-Fiber = 1237-52762-0533, B = 2.51 MG, S/N = 12 and T$_\mathrm{eff}$ = 8\,000~K; (b) SDSS J215148.31+125525.2, Plate-MJD-Fiber = 0733-52207-0522, B = 20.71~MG, S/N = 20 and T$_\mathrm{eff}$ = 10\,000~K; (c) SDSS J134845.98+110008.8, Plate-MJD-Fiber = 5445-55987-0530, B = 202.58~MG, S/N = 11 and T$_\mathrm{eff}$ = 16\,500~K. We corrected with a third degree polynomial function the inclination of the spectra and attributed an arbitrary shift in flux for a better visualization. The gray vertical lines represent the wavelength of the Balmer lines in the regime of no magnetic field.}
    \label{fig:example}
\end{figure}

\begin{figure}
    \centering
	\includegraphics[width=0.45\linewidth]{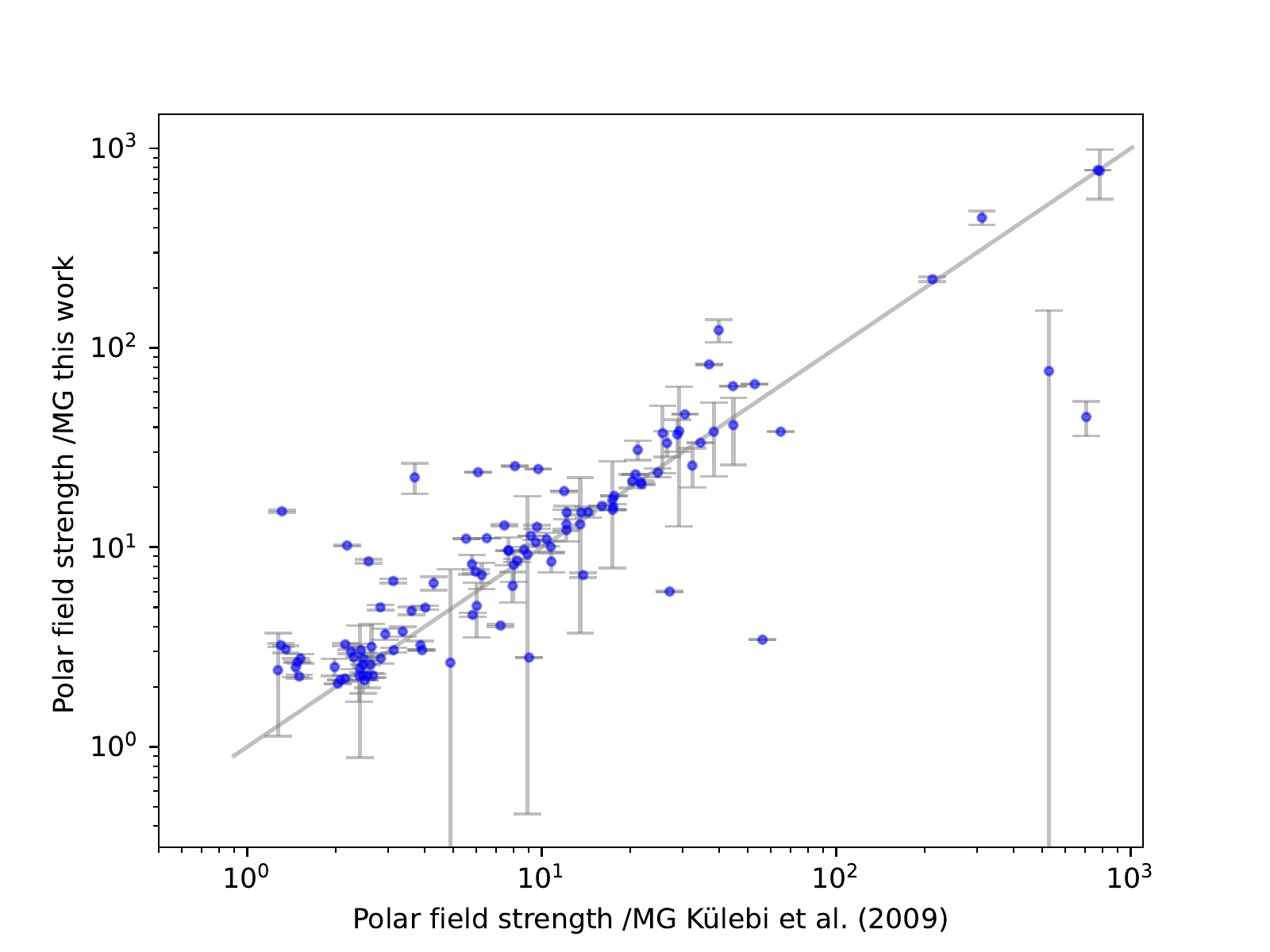}
        \includegraphics[width=0.45\linewidth]{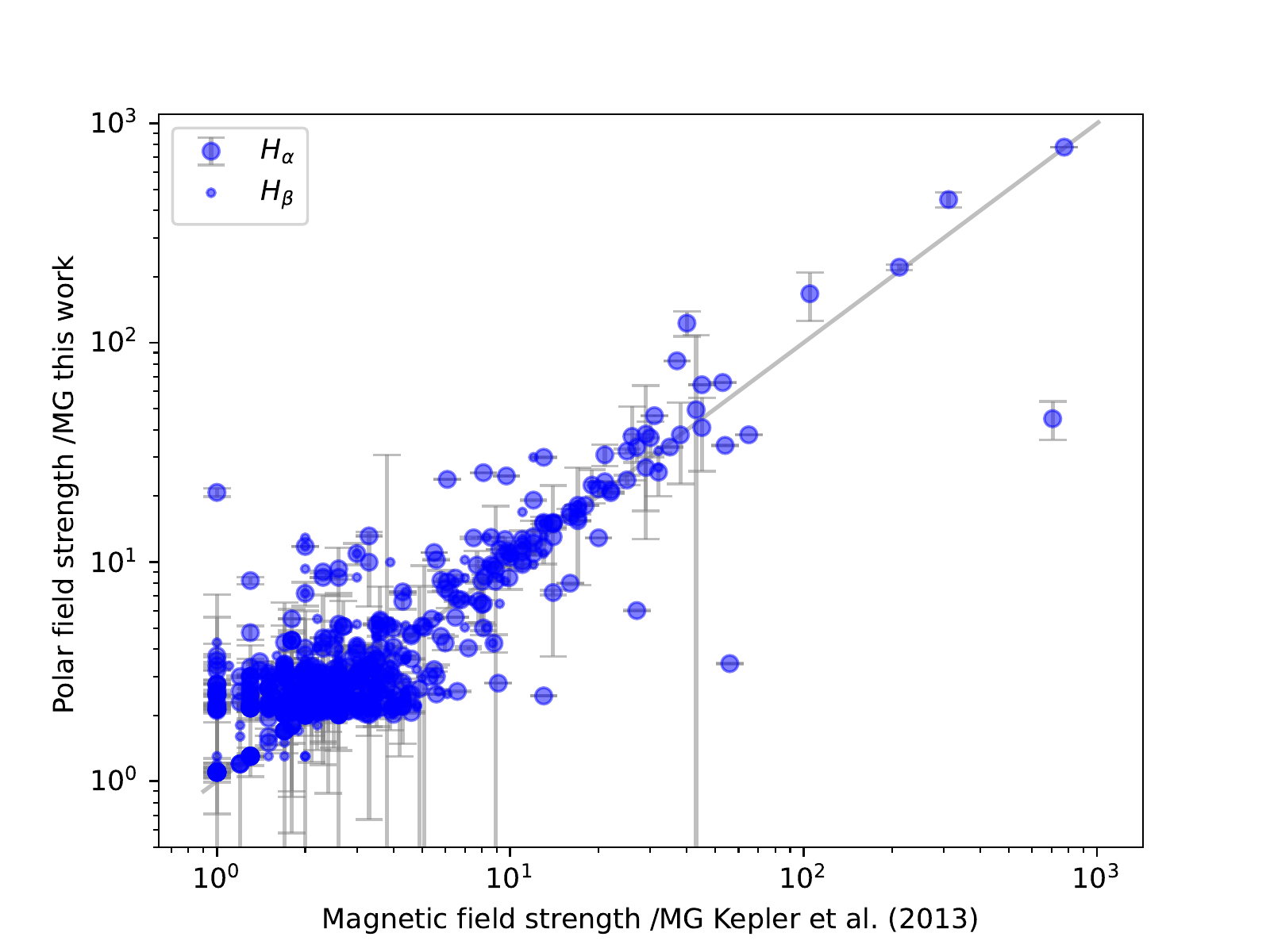}
    \caption{In the left panel, we present the comparison of dipole magnetic field strength of the offset dipole presented in this work and the values presented by \cite{2009Kulebi}. In the right panel, we present the comparison of dipole magnetic field strength of the offset dipole presented in this work and visually estimated values from \cite{2013Kepler}; the different size dots represent the two independent lines they used to estimate the magnetic field.}
    \label{fig:comp}
\end{figure}

Our results are presented in Table~\ref{tab}. To exemplify, Figure~\ref{fig:example} shows three white dwarfs with magnetic field strengths of different orders of magnitude. The values presented are compatible with previous determinations within the uncertainties.
We chose to use a fixed temperature extracted from the photometry of SDSS, while \cite{2009Kulebi} allowed it to be a free parameter in the fit. \cite{2013Kepler} only computed a visual estimate of the magnetic field considering only the spectral line positions. The comparison of these different methods is presented in Figure~\ref{fig:comp}.


The inclination and offset of the dipole are correlated quantities in our models. Different combinations between them can result in distributions of the magnetic field over the stellar surface for which the effect measured in one spectrum is the same. With that in mind, we interpret them as a measurement of the complexity of the field over the stellar surface and not necessarily a reflection of unique parameters. 

\begin{table}[]
    \centering
	\caption{The table presents the identifiers of the DAHs in the SDSS and their masses and effective temperatures calculated with Gaia astrometry presented in \cite{2021Gentile-Nicola-Gdr3} when available. $T_\mathrm{SDSS}$ is the temperature we used for the magnetic field model, computed using non-magnetic atmospheric models and SDSS colors g, r, and i. The following columns present the parameters of the magnetic field resulting from the best YAWP model for each star. In order, there is the  the dipole magnetic field strength of the offset dipole, the uncertainty computed through least squares, the z-offset from the center, and the inclination of the dipole to the line of sight. We call the attention that these last two quantities should be interpreted as a reference for the magnetic field structure (how different from a regular dipole) because there are degeneracies, especially due to changes in the field structure as the star rotates. The full table is available as online supporting information. 
	}
	\label{tab}
    \begin{tabular}{lrrrrrrrr}
    SDSS          	&	Plate-MJD-Fiber         	&	M (\Msun) &	$\mathrm{T}_\mathrm{eff}$ (K)  	 &	$\mathrm{T}_\mathrm{SDSS}$ (K)	&	B (MG)   	&	$\sigma$(B) (MG)  	&	$\mathrm{z}_\mathrm{offset}$ ($\mathrm{R}_*$) 	&	incl(°)\\
    \hline
    																	
J113212.99-003036.8	&	0282-51658-0278	&	---	&	---	&	20000	&	3.19	&	0.46	&	0.04	&	21.79	\\
J114720.40-002405.7	&	0283-51584-0120	&	0.68	&	15919	&	17000	&	2.00	&	---	&	---	&	33.47	\\
J121105.25-004628.5	&	0287-52023-0253	&	1.34	&	27473	&	19500	&	2.56	&	0.01	&	-0.32	&	66.65	\\
J121635.36-002656.3	&	0288-52000-0276	&	---	&	---	&	15000	&	64.24	&	0.11	&	-0.18	&	59.27	\\
J130807.48-010117.0	&	0294-51986-0089	&	0.47	&	11757	&	18500	&	2.20	&	0.00	&	-0.21	&	72.94	\\
J144114.21+003702.3	&	0307-51663-0595	&	0.87	&	25472	&	21000	&	4.39	&	0.49	&	0.49	&	20.70	\\
J112852.88-010540.7	&	0326-52375-0565	&	---	&	---	&	18000	&	2.81	&	0.36	&	0.36	&	0.08	\\
J113431.97-031529.0	&	0327-52294-0131	&	1.24	&	14539	&	15500	&	3.00	&	0.01	&	0.34	&	16.44	\\
J155238.20+003910.4	&	0342-51691-0639	&	1.26	&	14285	&	12500	&	2.16	&	0.08	&	0.12	&	41.32	\\
J171556.26+600643.7	&	0354-51792-0318	&	0.55	&	13567	&	9500	&	2.07	&	0.01	&	0.17	&	27.63	\\
J172932.48+563204.1	&	0358-51818-0239	&	---	&	---	&	10000	&	6.00	&	0.04	&	0.49	&	0.46	\\
J172329.14+540755.7	&	0359-51821-0415	&	1.09	&	10175	&	10000	&	36.82	&	6.75	&	-0.05	&	53.78	\\
J173915.64+545059.1	&	0360-51816-0547	&	0.30	&	9376	&	16000	&	2.69	&	0.02	&	0.37	&	31.91	\\
J173235.19+590533.3	&	0366-52017-0591	&	0.57	&	11028	&	11500	&	2.56	&	0.04	&	-0.36	&	78.63	\\
J171441.07+552711.3	&	0367-51997-0318	&	---	&	---	&	30000	&	6.45	&	1.80	&	0.47	&	3.84	\\
J172045.35+561214.8	&	0367-51997-0461	&	---	&	---	&	15000	&	24.67	&	0.01	&	-0.30	&	78.71	\\
J220435.05+001242.9	&	0372-52173-0626	&	1.11	&	10502	&	10500	&	2.15	&	0.00	&	0.13	&	12.83	\\
J220823.65-011534.1	&	0373-51788-0086	&	0.39	&	19044	&	9500	&	3.04	&	0.11	&	-0.50	&	35.80	\\
J220514.08-005841.6	&	0373-51788-0243	&	0.75	&	18464	&	14500	&	3.01	&	0.02	&	0.42	&	74.80	\\
J221828.58-000012.1	&	0374-51791-0583	&	1.06	&	12806	&	12239	&	220.78	&	6.35	&	-0.04	&	24.85	\\
J231432.89-011320.3	&	0382-51816-0289	&	---	&	---	&	18500	&	4.50	&	0.00	&	0.45	&	4.02	\\
J232248.21+003901.0	&	0383-51818-0421	&	1.17	&	18130	&	11500	&	21.20	&	0.49	&	-0.38	&	33.30	\\
J022335.15+004954.8	&	0406-51900-0490	&	0.54	&	6728	&	9500	&	2.16	&	0.02	&	-0.15	&	65.68	\\
J022523.67+002743.0	&	0406-51900-0543	&	---	&	---	&	15500	&	2.02	&	0.00	&	0.16	&	0.16	\\
J022623.80-002313.1	&	0406-52238-0071	&	---	&	---	&	16500	&	1.30	&	0.12	&	0.00	&	33.47   \\
J025837.19+000019.2	&	0410-51877-0065	&	0.75	&	10561	&	10000	&	2.33	&	0.01	&	0.16	&	2.51	\\
J032137.43+010437.3	&	0413-51821-0578	&	0.89	&	21140	&	18000	&	2.86	&	0.00	&	0.05	&	16.68	\\
J031323.65-001659.9	&	0413-51929-0313	&	---	&	---	&	30000	&	7.30	&	0.00	&	0.42	&	0.00	\\
J033145.69+004516.9	&	0415-51879-0378	&	1.05	&	19230	&	12000	&	13.05	&	2.37	&	-0.42	&	48.36	\\
J033320.37+000720.6	&	0415-51879-0485	&	0.79	&	7498	&	9500	&	771.71	&	214.35	&	0.18	&	0.10	\\
J034511.10+003444.2	&	0416-51811-0590	&	0.57	&	7431	&	8000	&	2.50	&	0.27	&	-0.36	&	12.91	\\
J003111.75+134919.5	&	0417-51821-0084	&	---	&	---	&	21000	&	2.24	&	0.00	&	0.15	&	0.72	\\
J003232.07+153126.6	&	0418-51817-0346	&	0.35	&	9917	&	17000	&	2.00	&	0.78	&	0.10	&	15.36	\\
J004513.88+142248.1	&	0419-51879-0147	&	0.81	&	7629	&	8000	&	3.61	&	0.04	&	-0.29	&	18.68	\\
J013533.20+132249.8	&	0426-51882-0291	&	---	&	---	&	21000	&	5.10	&	0.01	&	0.00	&	33.47	\\
J013920.54+152218.7	&	0426-51882-0524	&	0.46	&	10530	&	18000	&	2.30	&	0.00	&	0.19	&	0.56	\\
J021230.00+122557.2	&	0428-51883-0046	&	---	&	---	&	15500	&	2.07	&	0.11	&	0.14	&	0.80	\\
J075959.57+433521.1	&	0437-51869-0369	&	---	&	---	&	9000	&	91.55	&	57.78	&	-0.38	&	29.35	\\
J081136.33+461156.4	&	0439-51877-0523	&	---	&	---	&	40000	&	4.95	&	2.77	&	0.49	&	3.90	\\
J085159.32+532540.3	&	0449-51900-0311	&	0.87	&	11106	&	11500	&	63.76	&	28.16	&	-0.17	&	13.81
\end{tabular}
\end{table}

The magnetic field affects the line profiles, and we cannot use them to estimate surface gravity directly. Thereby, the effect of the magnetic field in the radiation transfer in model atmospheres used in the code were calculated with $\log g = 8$ for all the stars as an approximation. This value was originally chosen to consider that it is the mean value for white dwarfs. We acknowledge that it is not the best value to represent magnetic white dwarfs once they were found to be more massive than non-magnetic ones. Fortunately, this doesn't affect the position of the lines that are closely related to the magnetic field strength, only their strength.

The code also presents the possibility of searching for the temperature spectroscopically, together with the magnetic field parameters. But, due to the uncertainties in the spectra and the surface gravity versus effective temperature correlation, we opted to use a previously estimated temperature acquired by comparing the observed fluxes with hydrogen-rich atmospheric models. These models lead to a table of colors in the bands u, g, r, i, and z used by SDSS for each temperature. One can compare these values, especially their differences (color index), to the values from the observed spectra, thus obtaining an estimated temperature. We chose not to use the z-band since it is observed a lot of noise in the corresponding wavelengths. The u-band was also discarded since it is expected to be the most affected by the magnetic splitting of the lines.

We only fitted temperatures between 8\,000~K and 40\,000~K when varying the magnetic field due to convergence problems. For only one star out of 4 which we should have used 50\,000~K we were able to find a model with this temperature. No model was found with appropriate temperatures for the 13 stars cooler 8\,000~K. This limits our capacity to model the magnetic field of these stars, as the models present deeper lines than those observed in the SDSS spectra, partially due to the log g=8.0 approximation.
The depth of the absorption lines does not have a large impact on the determination of the magnetic field amplitudes, which are predominantly determined by the wavelength displacement and presence of splittings.
This effect can be seen in the upper part of Figure~\ref{fig:example}, in which a WD with a temperature below 8\,000 K is presented. The best model, in red, clearly has lines deeper than the observed spectra but well represents the splitting on the H$_\alpha = 6\,565$ \AA~and H$_\beta = 4\,861$ \AA~lines. 

\begin{figure}[ht]
    \centering
	\includegraphics[width=0.75\textwidth]{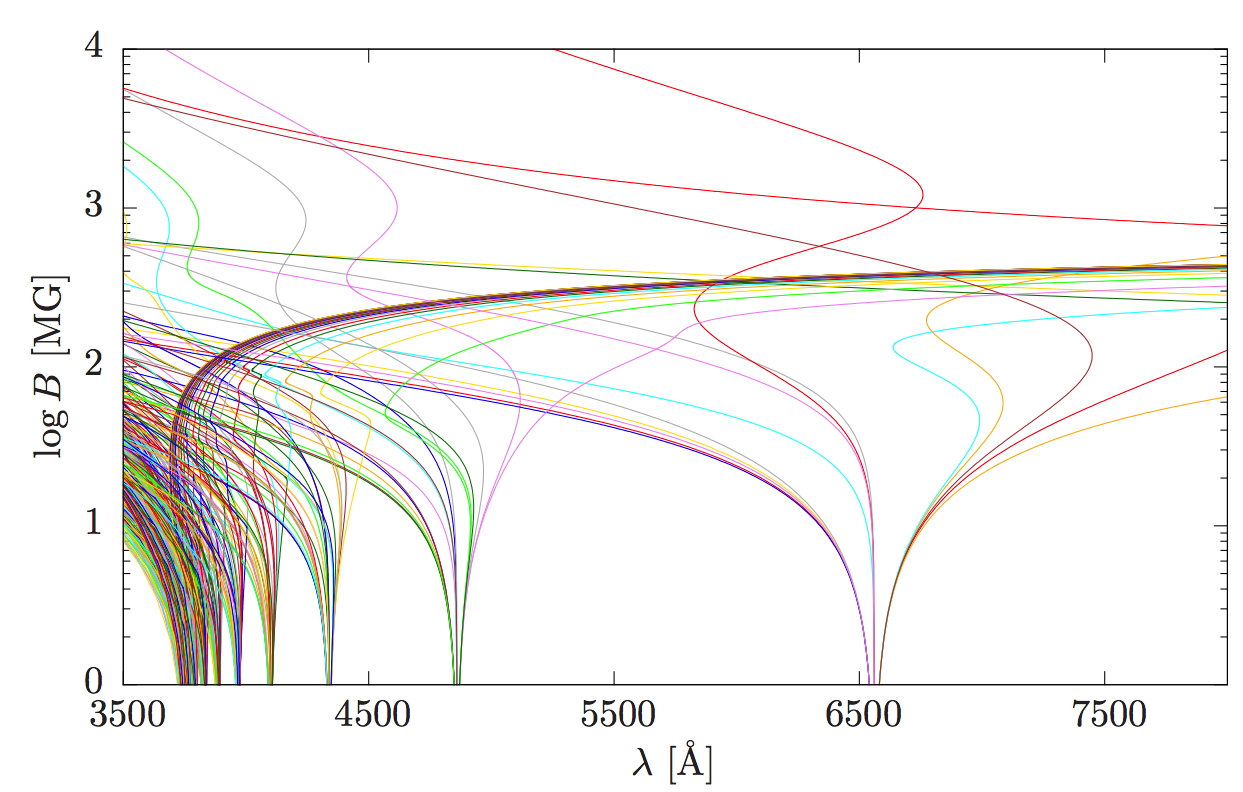}
    \caption{Magnetic field strength as a function of the wavelength of the first 325 transitions in the Balmer series, which emerge from the field-free Balmer transitions up to principal quantum numbers n~=~10. Colors were arbitrarily assigned to facilitate visualization. Figure from \cite{2014Schimeczek}}
    \label{fig:fields}
\end{figure}

For better visualization, we present the observed spectra with a running mean of five points because the measured spectra are usually noisy. We emphasize that this step was taken after we ran YAWP and its only purpose is to guide us in the analysis of the results. 
The noise in the observed spectra is wavelength-dependent and is significantly higher above 7\,000~\AA. It is noticeable that the hydrogen absorption lines do not reach these long wavelengths for field regimes below 25~MG ($\log B \approx 1.4$), as can be seen in Figure~\ref{fig:fields} from \cite{2014Schimeczek}. Therefore we do not consider higher wavelengths in the spectra of DAHs with lower fields to minimize the noise effect, which can be seen in the first two spectra of Figure~\ref{fig:example}. 

\begin{figure}[h]
    \includegraphics[width=\textwidth]{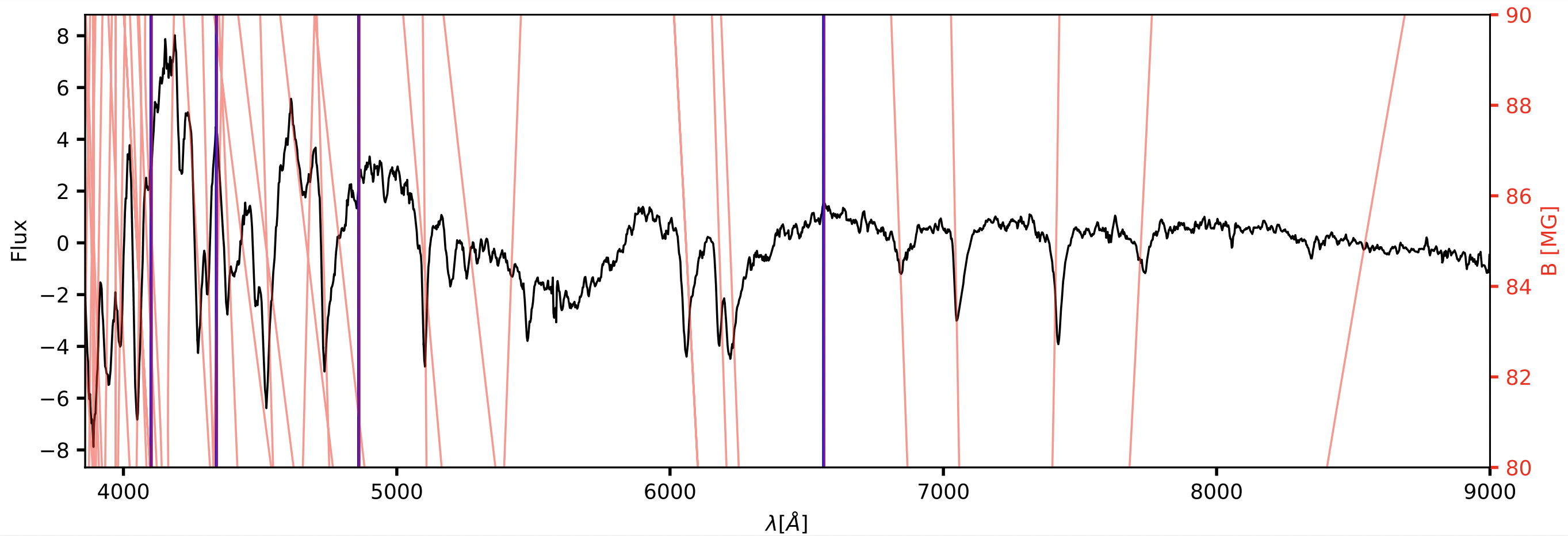}
    \caption{Spectra of DAH after applying the smooth function and normalizing by a third-degree polynomial function for better display.The star SDSSJ085649.67+253441.0 in the plot has Plate-MJD-Fiber = 5179-55957-0778, B = 85 MG, S/N = 14, and T$_\mathrm{eff}$ = 11\,500 K. The flux is $f_{\lambda}\ /\ 10^{-16}\ \rm{erg}\ \rm{cm}^{-2}\ \rm{s}^{-1}\ \AA^{-1}$. The red lines represent the magnetic field strength as a function of the wavelength, as computed by \cite{2014Schimeczek}. The blue vertical lines represent the position of absorption lines for hydrogen when no magnetic field is applied.}
    \label{fig:spagethivarias}
\end{figure}

For some stars, YAWP did not converge to a solution, 
and we resorted to a simpler visual analysis to estimate the magnetic field, used it as a fixed input to the code, and fitted only the inclination and the offset. An illustrative example of the visual inspection can be seen in Figure~\ref{fig:spagethivarias}. 

\begin{figure}[h]
\centering
    \includegraphics[width=0.75\textwidth]{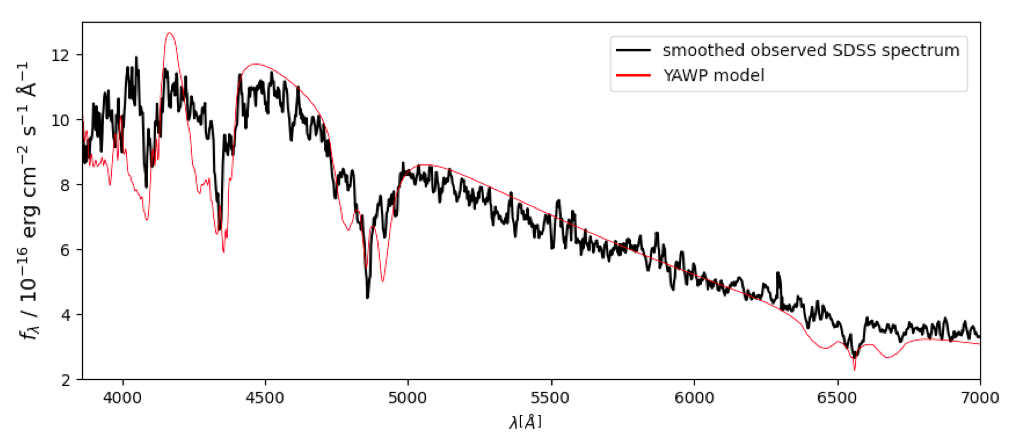}
    \caption{SDSS spectrum of the star with SDSS J004248.20+001955.2 is presented in black. The model with the best least-squares fit to the observed data with Plate-MJD-Fiber = 0690-52261-0594, B = 8.5~MG, S/N = 7.5 and T$_\mathrm{eff}$ = 14\,000~K is shown in red. Even though we plot the whole spectrum, the continuum is not used i the fit.}
    \label{fig:YAWP-b}
\end{figure}

After we determined a magnetic field strength visually, we computed the best model with the magnetic field fixed and varying only the inclination and offset. One example is portrayed in Figure~\ref{fig:YAWP-b}. In this specific case, YAWP did not converge to the best solution on its own, most probably due to the possible contamination of a very close star in the field.

\begin{figure}[h]
\centering
    \includegraphics[width=0.75\textwidth]{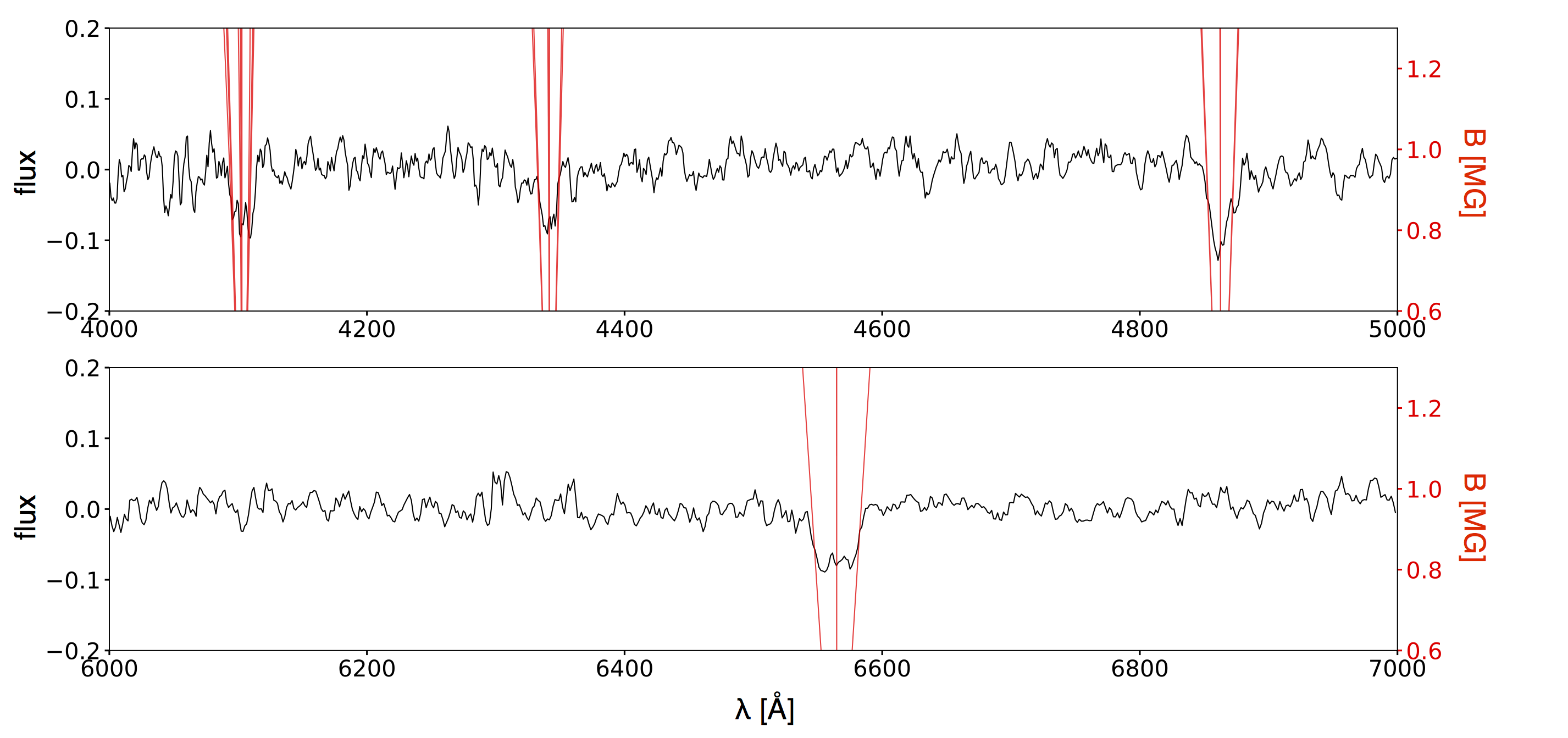}
    \caption{Spectra of the star with SDSS J090139.03+064022.4, Plate-MJD-Fiber = 4868-55895-0730, B = 0.8~MG, S/N = 12, and T$_\mathrm{eff}$ = 8\,000~K after applying the smooth function and normalizing by a third-degree polynomial function for better display. The flux is $f_{\lambda}\ /\ 10^{-16}\ \rm{erg}\ \rm{cm}^{-2}\ \rm{s}^{-1}\ \AA^{-1}$. The red lines represent the magnetic field strength as a function of the wavelength, as computed by \cite{2014Schimeczek}.}
    \label{fig:spagethi}
\end{figure}

The visual estimation was also used for four stars with magnetic fields below $1$~MG and S/N above 10. The magnetic field values were not used as input for YAWP since it is out of the range comprised in the models. An example can be seen in Figure~\ref{fig:spagethi}.

\begin{figure}[h]
    \centering
    \includegraphics[width=\textwidth]{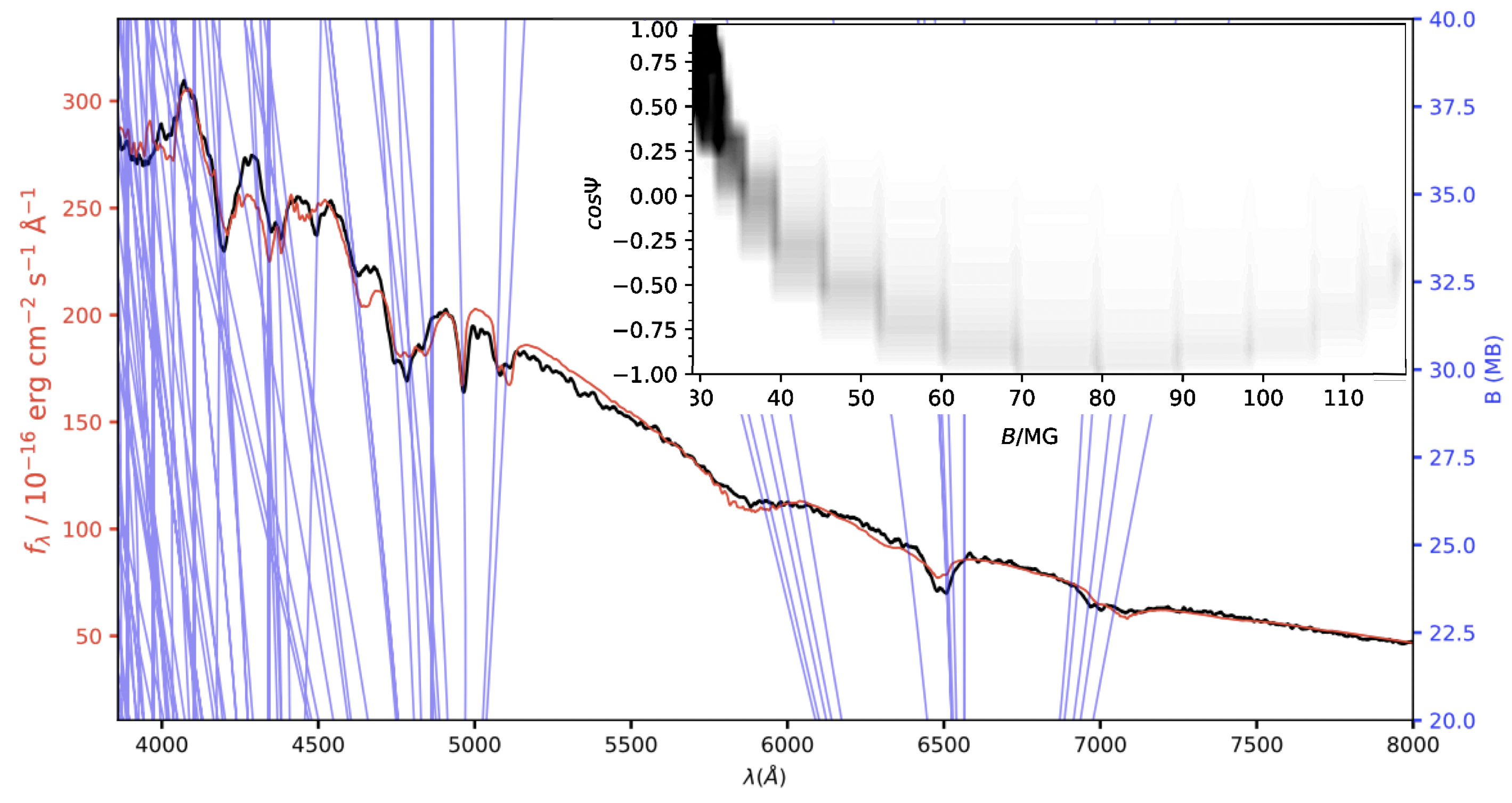}
    \caption{Outer panel showing the SDSS spectrum of the star with SDSSJ102054.10+362647.0, Plate-MJD-Fiber = 4568-55600-0952, B = 61.2 MG, S/N $\simeq$ 63 and $\mathrm{T}_\mathrm{eff}$ = 13\,500 K is presented in black. The model with the best least-squares fit to the observed data is shown in red. The blue lines represent the magnetic field strength as a function of the wavelength, as computed by \cite{2014Schimeczek}. The inner panel shows the distribution of magnetic field over the stellar surface due to the inclination of $29.35^\mathrm{o}\,$ and  z$_\mathrm{offset}$ of -0.38 \Rstar delineated by the best model.}
    \label{fig:YAWP-zebra}
\end{figure}

The only geometry of magnetic fields considered for our determinations was the non-centered dipole, which was a good approximation for these objects. Therefore, the difference in magnetic field strength between stellar surface regions is substantial, up to orders of magnitude, and is larger with increasing polar field strength. Some exceptions exist, such as if the inclination is close to $90\deg$, and we look straight into a pole of a little off-centered dipole. This effect can be seen in Figure~\ref{fig:YAWP-zebra}. The model that best reproduces the observed data has a mean modulus of magnetic field strength over the stellar surface of B = 61\,MG (red line in the outside panel). However, this is incompatible with simply summing the respective spectra of each area element and evaluating the result. Since each magnetic field intensity affects the line profiles in a specific way, the summed spectra will look more like the mode than the mean, in this case.

It can be seen in the inside panel of Figure 6 that about half of the stellar surface presents field intensities varying from 40 to 120\,MG, while the other half presents field intensities around 30\,MG. In addition, lower fields are easier to identify visually due to the lower complexity of the line profiles, as seen in Figure~\ref{fig:fields}. As a consequence, if we did the visual inspection of the total spectra coming from the whole visible disk, we would find a value closer to 
30\,MG.

In addition to the intrinsic variation, we estimate that our precision cannot be better than $1$\,MG due to spectral resolution and signal-to-noise ratios. Beyond these limitations, from the $\chi^2$, the mean magnetic field uncertainties are approximately 12\% of the computed value.

\section{Discussion}
\label{chap:Dis}

\cite{2003Liebert} propose that magnitude-limited samples, as the one studied in this work, have a bias against higher mass white dwarfs since for a given temperature, they have a smaller radius 
and consequently smaller luminosity. This would lead them to be detected less frequently. However, they did not take into consideration that more massive white dwarfs take longer to cool down because they have a smaller radius.
It is not straightforward which of these effects predominates at a given time. \cite{2021Bagnulo} showed that magnitude-limited samples have a complex bias against or in favor of higher masses depending on the stellar age. One can conclude that there is a bias in favor of younger stars independently of their mass. 
They also found that the frequency of magnetic white dwarfs is substantially depressed for stars younger than 0.5\,Gyr and that this difference probably reflects the action of the mechanisms that produce magnetic fields in white dwarfs.

Even though it favors younger stars, magnitude-limited surveys, especially with low-resolution spectroscopy, are the ones that can go deeper in magnitude and examine a larger sample of stars. With this perspective, we will discuss the distribution of magnetic field strength, the relation between magnetic field and mass, effective temperature and period, and some specific cases that stand out in our sample.

\subsection{Distribution of magnetic field strength}

The fraction of magnetic white dwarfs rich in hydrogen found in this work was 2.7\%, far below the previous values presented in the literature. We call attention to the strong bias present in our sample due to the chosen survey. The selection of which stars were observed by SDSS changed over time, resulting in a smaller number of magnetic white dwarfs in the latest data releases because white dwarfs were not specifically targeted.

\begin{figure}[h]
    \centering
    \includegraphics[width=0.75\textwidth]{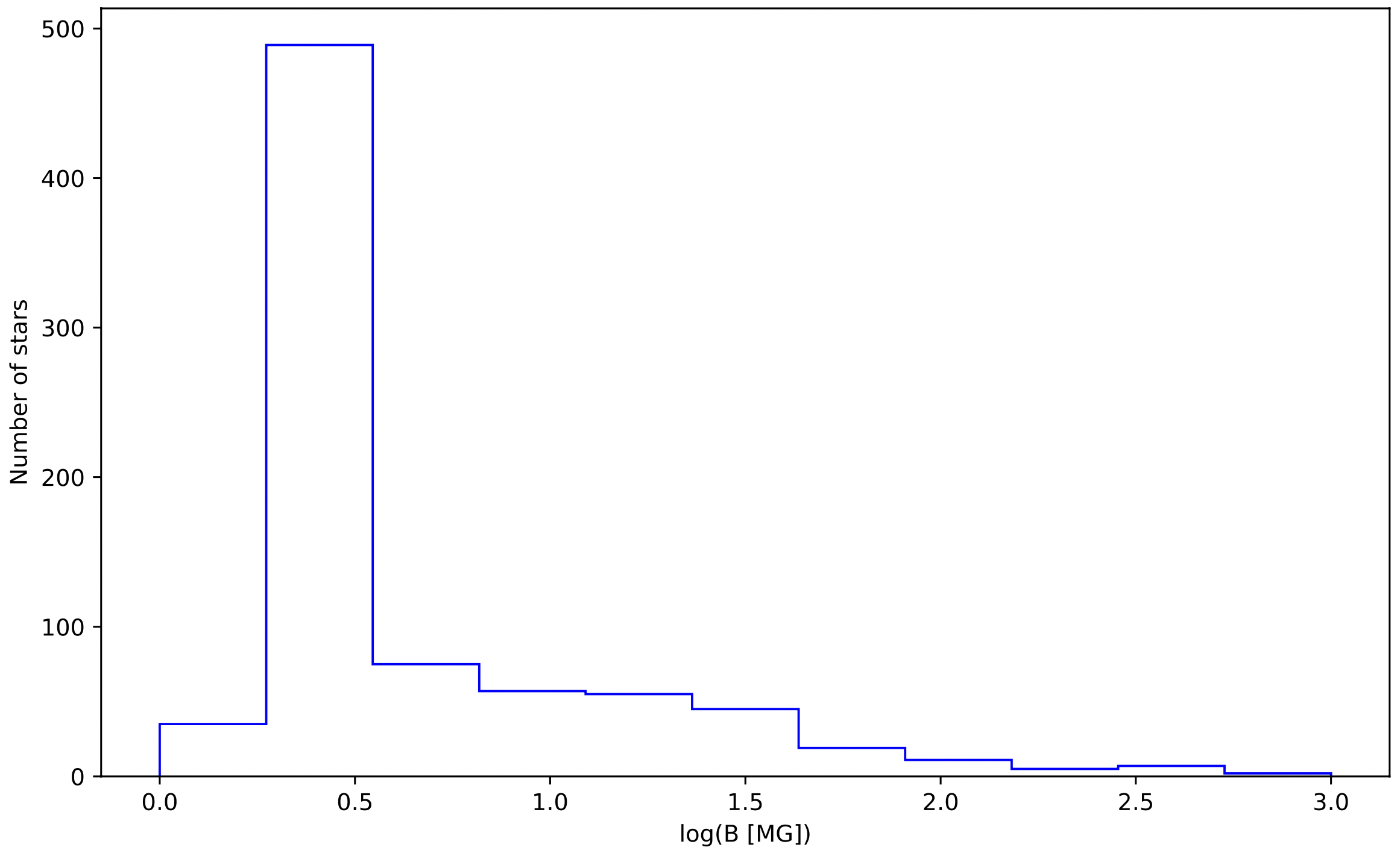}
    \caption{Magnetic field strength histogram for all magnetic white dwarf in our sample.}
    \label{fig:hist_logb}
\end{figure}

We also note that the visual identification of Zeeman splittings is much more efficient for magnetic fields below 60~MG when the effect is well-behaved, as can be noticed in Figure~\ref{fig:fields}. The SDSS spectral resolution around 2~\AA~also hinders the detection of fields below 1~MG, even for S/N$\geq$10.  With these reservations, we present the distribution of the magnetic field strength of our sample in Figure~\ref{fig:hist_logb}. A higher appearance of magnetic fields strength below 3\,MG is noticeable. It is not in all cases evident that the stars are magnetic due to the limited SDSS S/N.

\subsection{Relation between mass and magnetic field}
\begin{figure}[h]
    \centering
    \includegraphics[width=0.8\textwidth]{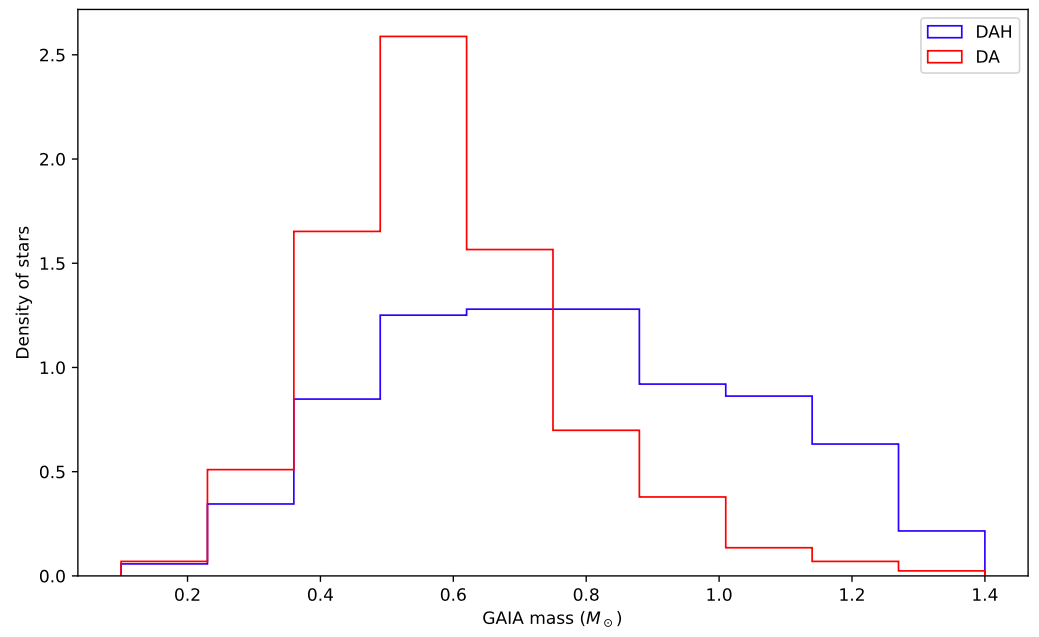}
    \caption{Histogram of mass calculated with Gaia data for all DAs in red and only the magnetic ones in blue.}
    \label{fig:hist_mass}
\end{figure}

\begin{figure}[h]
    \centering
    \includegraphics[width=0.495\textwidth]{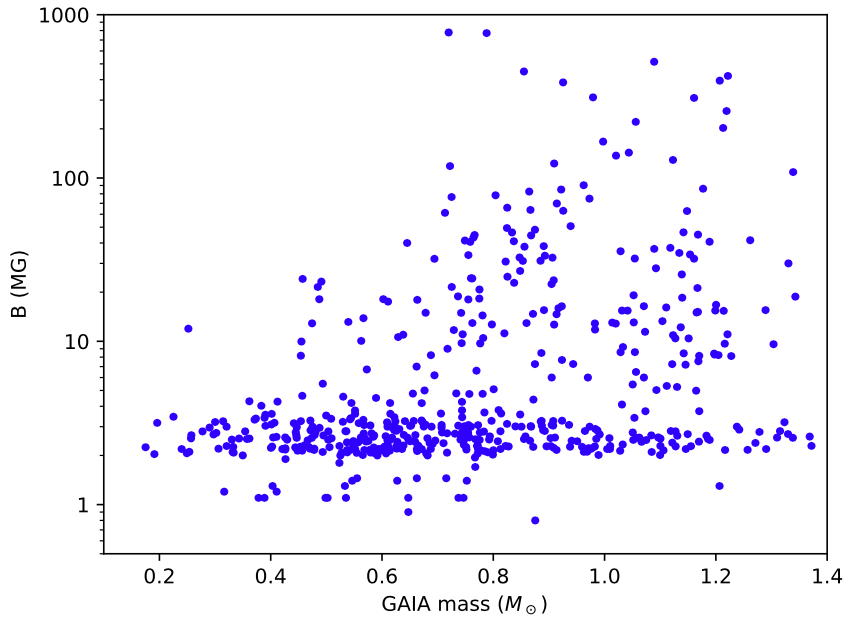}
    \includegraphics[width=0.495\textwidth]{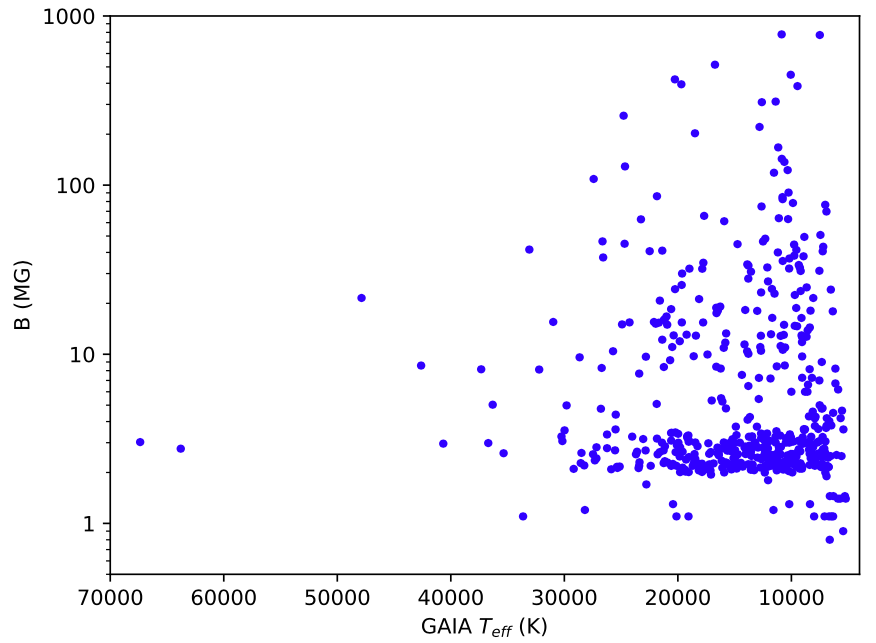}
    \caption{Magnetic field versus mass calculated with Gaia data, showing a clear absence of highly magnetic white dwarfs with lower masses on the left. Magnetic field versus effective temperature calculated with Gaia data, showing a clear increase of highly magnetic white dwarfs with lower temperatures on the right.}
    \label{fig:mass_B}
\end{figure}

To examine the possible relation between magnetic field and mass, we made the Figure~\ref{fig:hist_mass} to study the fraction of stars with magnetic fields, and Figure~\ref{fig:mass_B} to study the connection between the strength of the magnetic field and the star's mass.

It appears that the distribution of mass of DAHs can be approximated by a  Gaussian centered in 0.78\,\Msun\ with no further remarkable features.
The left panel in Figure~\ref{fig:mass_B} shows an evident lack of low mass DAHs with strong fields.

\subsection{Relation between effective temperature and magnetic field}
\label{sec:tef}

In the search for a hint of the magnetic field origin, we investigate its relation to the effective temperature of the inspected stars. The right panel in Figure~\ref{fig:mass_B} shows that the magnetic field strength increases as the effective temperature decreases.

\begin{figure}[h]
    \centering
    \includegraphics[width=0.75\textwidth]{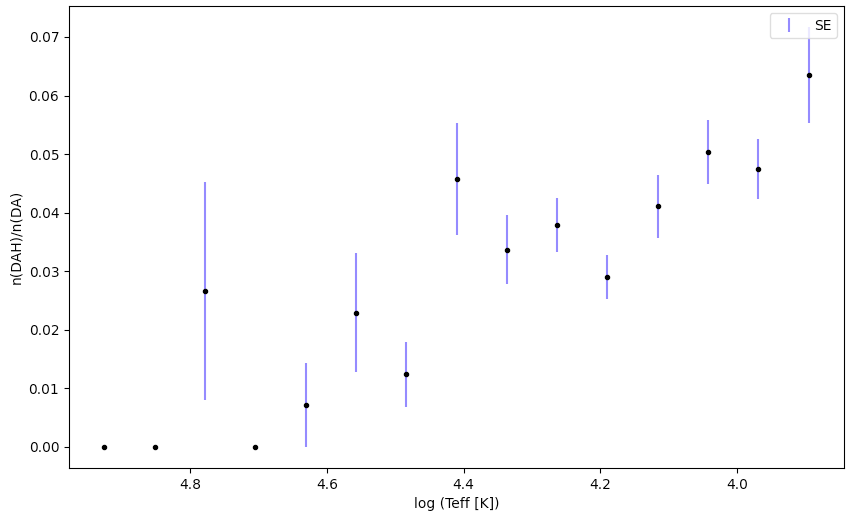}
    \caption{The ratio between the number of DAHs and the number of DAs versus effective temperature, showing a peak of abundance near T$_\mathrm{eff} = 25\,000$ K. The blue vertical lines represent the standard errors for each point.}
    \label{fig:da/dah}
\end{figure}

Not only do we find white dwarfs with stronger magnetic fields, but also we detect more magnetic white dwarfs at lower temperatures. The second effect is naturally expected because it is easier to detect white dwarfs at lower temperatures since they spend more time cooling down. To consider this effect and check if there are really more magnetic white dwarfs at lower temperatures, we study the fraction of DAHs compared to the whole sample of white dwarfs rich in hydrogen. This is illustrated by Figure~\ref{fig:da/dah}, and it is noticeable that the fraction of magnetic stars is indeed increasing as the temperature decreases.

It is especially outstanding the rapid growth in magnetic fraction around effective temperature of $25\,000$\,K ($\log T_\mathrm{eff} \approx 4.4$), the temperature at which a convective zone of helium is internally formed in the white dwarf, that could be responsible for this increase. We also call attention to the build-up in the magnetic fraction that starts around  the effective temperature of $16\,000$\,K ($\log T_\mathrm{eff} \approx 4.2$). At this temperature, a convective zone of hydrogen is formed in the white dwarf, and it could be responsible for the rise in the magnetic fraction.

\begin{figure}[h]
    \centering
    \includegraphics[width=0.495\textwidth]{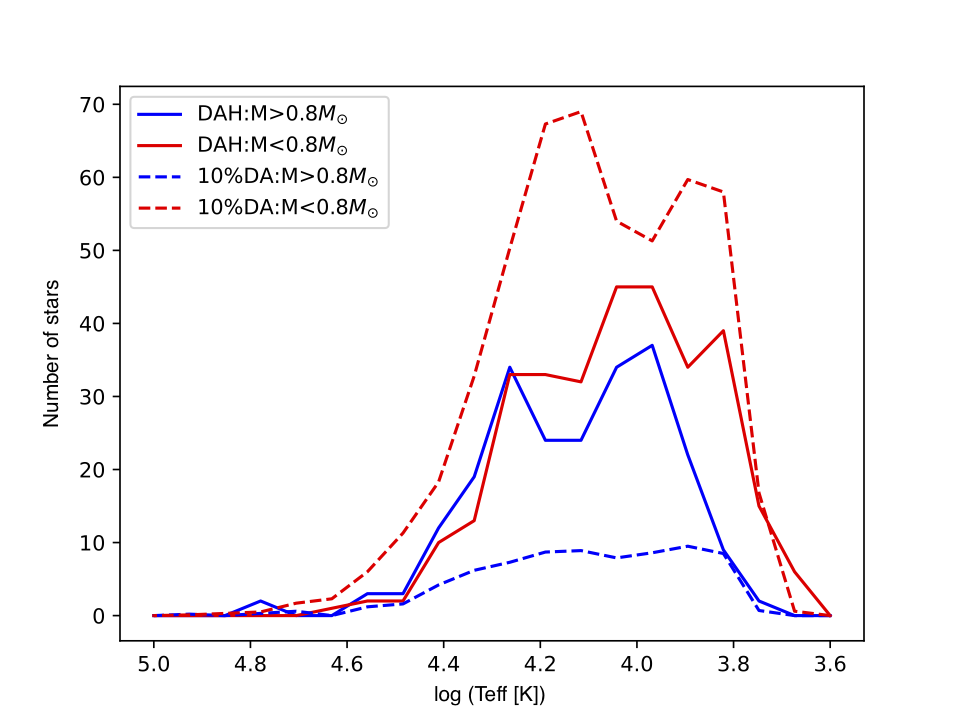}
    \includegraphics[width=0.495\textwidth]{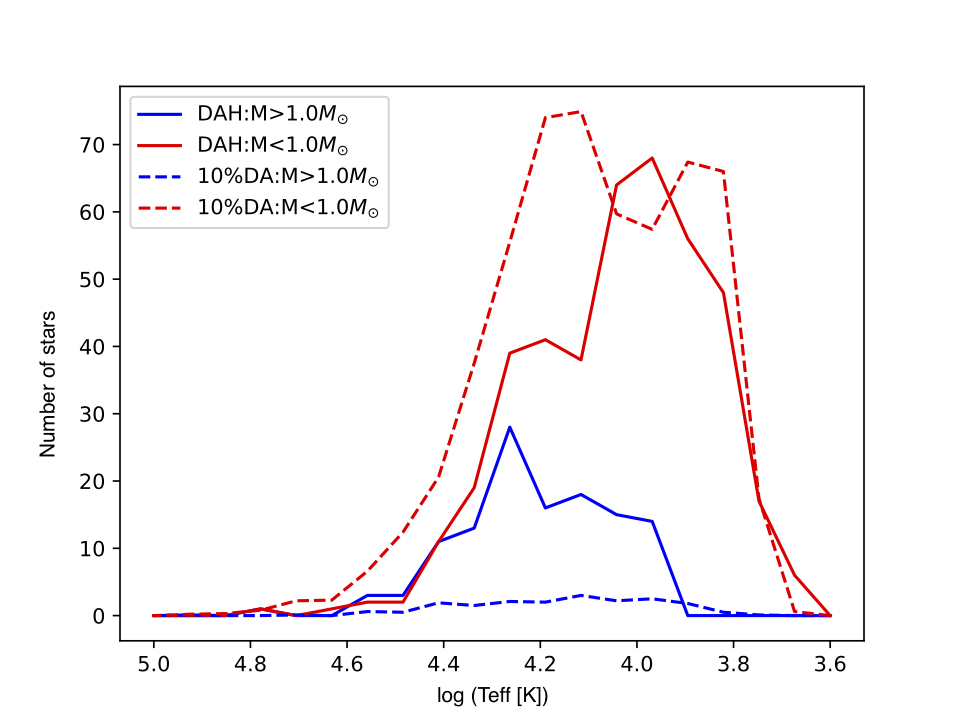}
    \caption{Both panels show the number of DAHs and DAs versus the effective temperature. The sample was divided at 0.8 \Msun and 1.0 \Msun, represented in the upper panel and the lower panel respectively. We divided the absolute number of DAs by ten to allow better visualization. It is evident a change in behavior of the distribution of DAHs that is not accompanied by the distribution of DAs.}
    \label{fig:da/dah_mass}
\end{figure}

A question then arises: does the effect change depending on the stellar mass? To answer this question, we divided our sample in two at M = 0.8\,\Msun~and at M = 1\,\Msun~and compared the results, which can be seen in the left and right panel of Figure~\ref{fig:da/dah_mass} respectively.

The distribution with temperature of DAs doesn't change much, except for the fact that there are considerably fewer stars with M $>$ 1 \Msun. Differently, for the DAHs, there is a significant variation of the distribution. We highlight the valley around $10\,000$~K (log T$_{eff}$ = 4.0) as  being a consequence of convective mixing and dilution. They pollute the stellar atmosphere with helium and thin the hydrogen layer, reducing the number of DAs and increasing the number of DABs or DBAs. This effect is constrained in temperature as a consequence of the disappearance of the lines of helium at lower temperatures, meaning that the star may contain helium in its atmosphere, it is only not possible to observe it through the spectra.

The magnetic DAs with mass above 0.8 \Msun have a similar double peak behavior, even though it is more restricted in temperature. The magnetic DAs with mass above 1 \Msun~on the other hand have the second peak missing. Some mechanism must be inhibiting the magnetic field for higher masses at temperatures around $16\,000$ K ($\log T_{eff} \approx 4.2$). This same effect is not observed for masses below 1 \Msun. In fact, the opposite occurs, and they show the second peak much higher than the first. The behavior of the lower mass DAHs is precisely the one discussed earlier, thus becoming necessary the understanding of what could be suppressing the higher masses magnetic fields.It is valid to remember that more massive stars are usually less detected in magnitude limited samples because they are fainter, so this could also be the result of selection effects.
\subsection{Relation between crystallization and magnetic field}
\label{sec:crys}

\begin{figure}[h]
    \centering
    \includegraphics[width=0.67\textwidth]{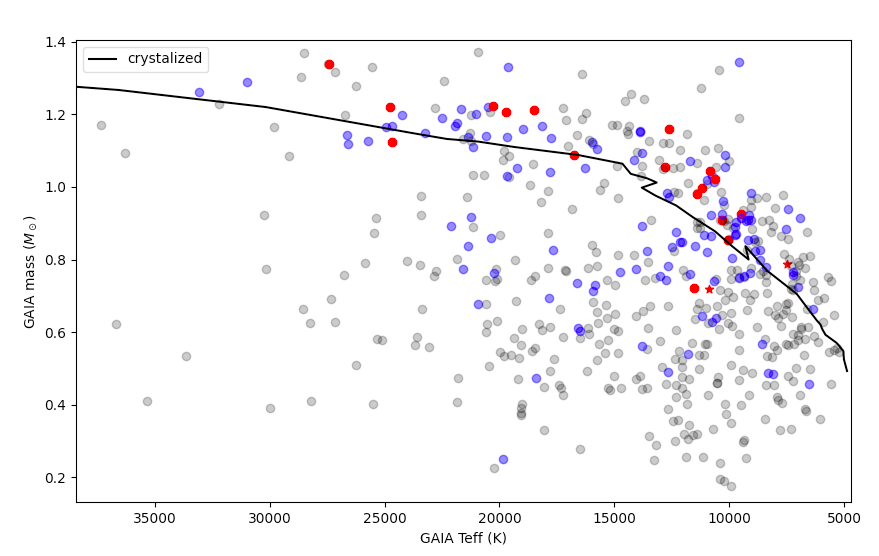}
    \caption{Mass versus effective temperature calculated with Gaia data. The colors black, blue, and red represent magnetic fields below 10\,MG, between 10---100\,MG, and above 100\,MG, respectively. The two red stars represent the most magnetic DAs with a magnetic field above 700\,MG. The solid black line represents the temperature of crystallization for the single evolution of white dwarfs of different masses as presented by  \cite{2013Romero},  \cite{2010Horowitz}, and  \cite{2018Lauffer}. Crystallization increases to the right from the line, i.e., cooler temperatures.}
    \label{fig:crist3}
\end{figure}

One significant physical process that is highly dependent on the white dwarf mass is the crystallization of its core, and we suppose that it is the one holding the magnetic field back. In the search for a better understanding of the relation between the crystallization and the magnetic field, 
we built Figure~\ref{fig:crist3}, in which one can see not only if the star has started to crystallize its core but also have an idea of how advanced this process is. The further the star is from the crystallization line in the colder direction, the more crystallized its core is. We emphasize that the crystallization line has an intrinsic uncertainty to the models used to compute it. 

It is recognizable that most of the stars with higher magnetic fields have started the crystallization process, which goes against the hypothesis that crystallization is responsible for the lower fraction of magnetic DAs with higher masses. \cite{2017Isern} proposed a mechanism of generation of magnetic fields of strengths of up to 0.1 MG by the dynamo in the convective region generated by the phase separation due to the crystallization process.  \cite{2022Ginzburg} found that these fields could go as high as 100 MG depending on the rotational period of the star and its mass.
\subsection{Particular Stars}
\begin{figure}[h]
\centering
	\includegraphics[width=0.82\textwidth]{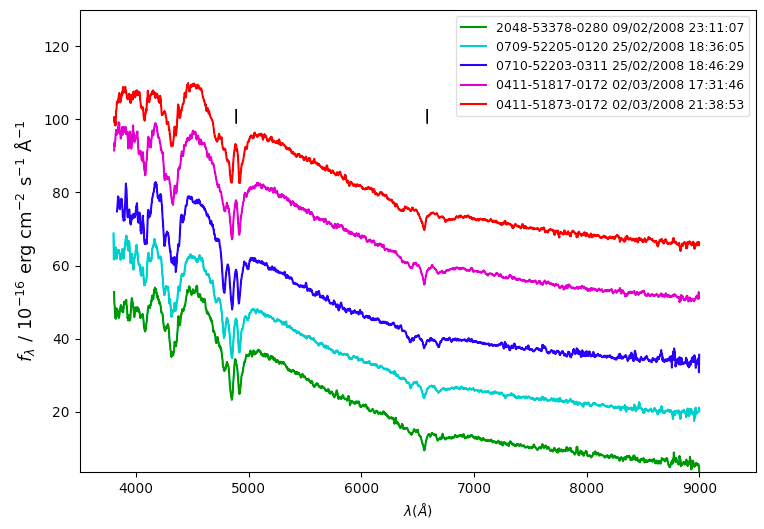}
    \caption{Different spectra of SDSS J030407.40-002541.74 with their respective observation dates. The two small vertical black segments mark the position of H$_\beta$ and H$_\alpha$ lines.}
    \label{fig:various}
\end{figure}

Some stars have more than one spectrum observed by SDSS in which it is possible to look for variability in the line profiles as an indication of the rotational period. Besides the inclination of the star in respect to the line of sight, which may allow us to see different portions of the stellar surface as it rotates, the line profiles can also change due to the misalignment between the magnetic field axis and the rotation axis. Here we highlight the star SDSS~J030407.40-002541.74 in which this effect is prominent. In Figure~\ref{fig:various} we can see that the $H_\alpha$ line varies between a deeper central line to equally deep triplet components. It is also visible that the $\sigma-$ component of the $H_\beta$ line appears and disappears.

This star has also been observed by TESS, and we have 30 minutes cadence data from sectors 4 and 31. Unfortunately, no variation above the detection limit was identified, which is understandable since it is very faint (Gaia Mag = 17.8528).

\begin{figure}[h]
    \centering
    \includegraphics[width=0.95\textwidth]{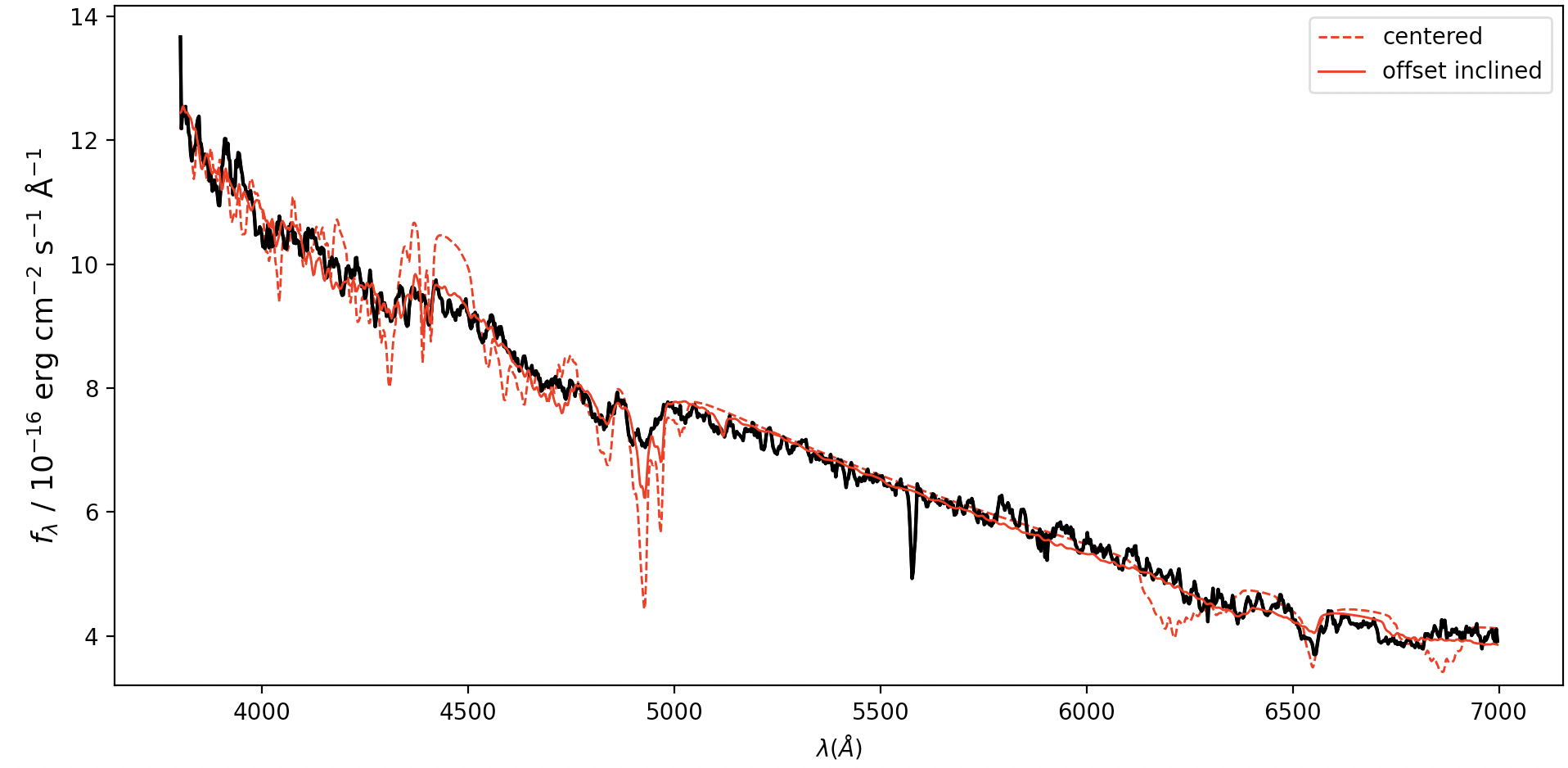}
    \caption{Observed spectrum of J221141.80+113604.5 with Plate-MJD-Fiber = 5064-55864-0122, B=18.77\,MG, S/N=20 and $T_\mathrm{eff} = 9\,500$\,K. The model with the best least-squares fit to the data is shown in red. We highlight that the line near $5\,500$~\AA~is due to a systematic error of joining two different observations by SDSS. The full red line shows a model with computed offset magnetic dipole field geometry, while the dashed red line shows a model computed not inclined centered magnetic dipole field geometry with the strength found in the previous calculation.}
    \label{fig:kilic-centered}
\end{figure}

The same process that affects the SDSS~J030407.40-002541.74 $H_\beta$ lines could be responsible for the $H_\beta$ line profile of SDSS~J221141.80+113604.5. which is illustrated in Figure~\ref{fig:kilic-centered}.  \cite{2021Kilic} argued that assuming a hydrogen atmosphere, an inclination, and an offset dipole geometry, the lines could not be reasonably reproduced. We found a good fit to the observed spectra except for the lateral components of the $H_\beta$ line, which is expected to change as shown for SDSS~J030407.40-002541.74.

\cite{2021Kilic} assumed a centered dipole as in 
Fig.~\ref{fig:kilic-centered} lower panel, and explained the difference in the line depth between the model and the observed spectra with the assumption of a non-pure hydrogen atmosphere. Our model does not require this additional complexity to explain the data.

\begin{figure}[h]
    \centering
    \includegraphics[width=0.75\textwidth]{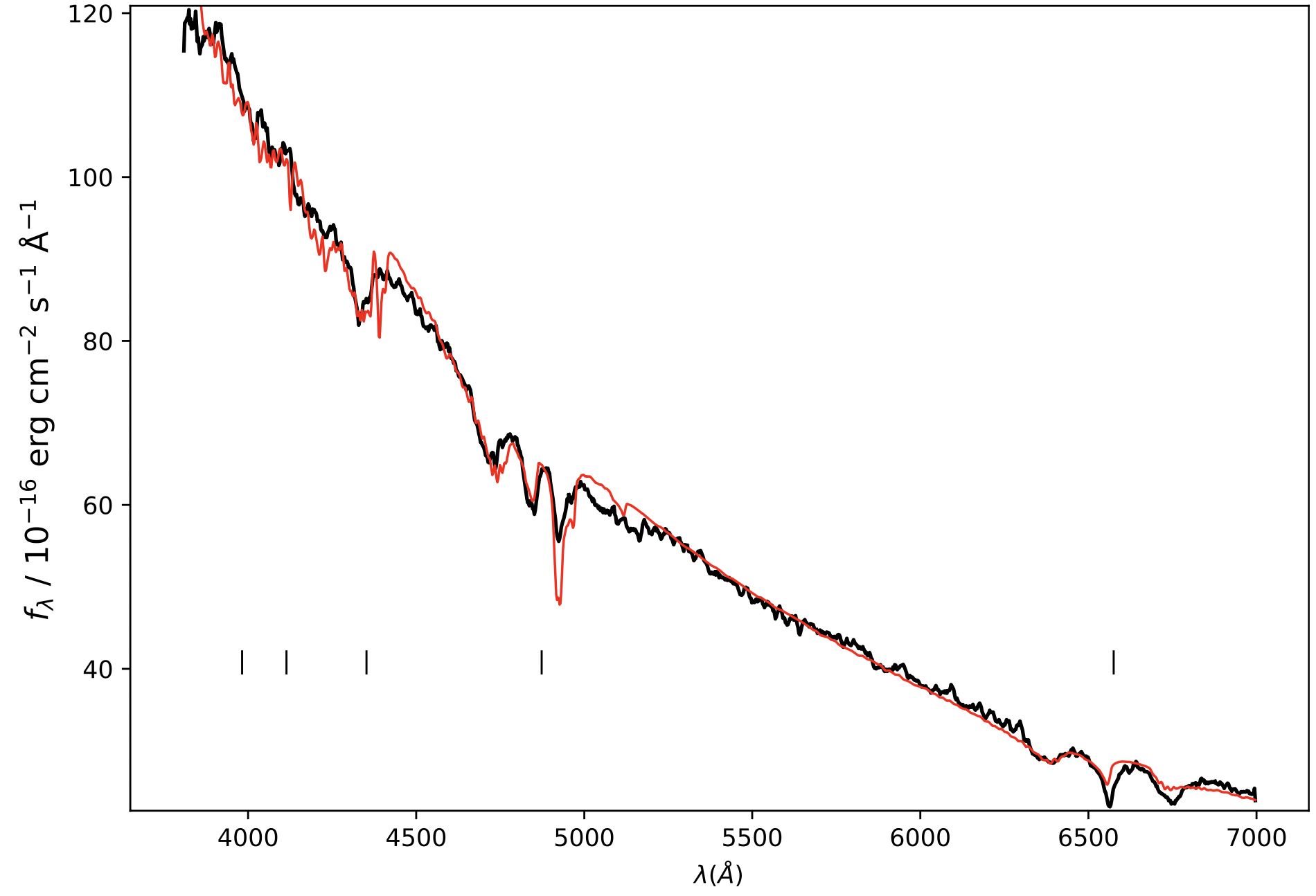}
    \caption{Observed spectrum of SDSS~J225726.05+075541.6 with Plate-MJD-Fiber = 2310-53710-0420, B=17.4\,MG, S/N=34 and $T_\mathrm{eff} = 9\,500$\,K. The model with the best least-squares fit to the data is shown in red.}
    \label{fig:francois}
\end{figure}

Another interesting case is the star SDSS~J225726.05+075541.6 featured in \cite{2022Williams}. They stated that the observed Balmer lines are significantly weaker than predicted by the atmospheric models, which could be explained if the star were in an unresolved binary system, but the allowable parameter space for such a binary is minuscule. We found a good fit, again with only one component of $H_\beta$ with large discrepancies. Our model is shown in Figure~\ref{fig:francois}.

\section{Conclusions}
\label{chap:conc}
In this work, we estimated magnetic field strength for 804 white dwarfs observed from SDSS of which 287 are new discoveries. Save rare exceptions, our determinations are coherent with \cite{2013Kepler} and \cite{2009Kulebi}. It consists of the largest number of magnetic field determinations for white dwarfs to date. We searched for relations between magnetic field strength and stellar mass, effective temperature, and crystallization status.

It was found that a considerable percentage of DAH have fields below 3 MG. This is, to some extent, biased by our spectroscopic method, but could also mean that lower fields are more abundant in white dwarfs. This result is in opposition to  \cite{2021Bagnulo} which concluded that within the range of field strength found in the 20\,pc volume, which extends between about 40\,kG and 300\,MG, the probability of fields occurring is roughly constant per dex of field strength.

We found that the magnetic field strength increases as the effective temperature decreases, together with an increase in the fraction of magnetic white dwarfs. This effect corroborates with the surface magnetic field being generated or enhanced in the white dwarf cooling phase. 
We could also observe that the highest fields tend to occur in the more massive stars and that the mean mass was, in general, higher than the non-magnetic ones ($0.78\,M_\odot$ compared to the mean $0.6\,M_\odot$). This does not give us any new information about the origin of the magnetic field because many assumptions already consider a higher mass. But a physical property closely related to the mass is crystallization, and we found that the most magnetic ones tend to be already crystallized. This is expected since DAH have higher masses, so they crystallize at higher temperatures but spend more time (easier to detect) at lower temperatures (already crystallized).

The general behavior found in this work is compatible with the findings by \cite{2022BagnuloLandstreet} that a high fraction of high-mass WDs have a strong magnetic field very early in their cooling phase, while normal mass stars are rarely magnetic when they are hot (young), but when they get cooler (older), magnetic fields become more common and stronger with time. 
Some other mechanism may be inhibiting magnetic fields in WDs with $M\geq 1\,M_\odot$  and $T_\mathrm{eff}\leq 16\,000$\,K.




\begin{acknowledgments}
We thank the support by grants from CAPES and CNPq (Brazil).
\end{acknowledgments}

%

\vspace{5mm}
\facilities{SDSS \cite{2021Kepler}, Gaia \cite{2021Gentile-Nicola-Gdr3}.}


\software{astropy \citep{2013A&A...558A..33A,2018AJ....156..123A},  
          YAWP \cite{2009Kulebi}.}



\appendix

\section{Complete table}
\startlongtable
\begin{deluxetable}{lrrrrrrrr}
\tablecaption{The table presents the identifiers of the DAHs in the SDSS and their masses and effective temperatures calculated with Gaia astrometry presented in \cite{2021Gentile-Nicola-Gdr3} when available. Then we have the temperature used for the magnetic field model, computed using non-magnetic atmospheric models and SDSS colors g, r, and i. The following columns present the parameters of the magnetic field resulting from the best YAWP model for each star. In order, there is the dipole magnetic field strength of the offset dipole, the uncertainty computed through least squares, the z-offset from the center, and the inclination of the dipole to the line of sight. We call attention that these last two quantities should be interpreted as a reference for the magnetic field structure because there are many degeneracies for the geometry. \label{tab}}
\tablehead{\colhead{SDSS} &	Plate-MJD-Fiber & M & $\mathrm{T}_\mathrm{eff}$ & $\mathrm{T}_\mathrm{SDSS}$ & B & $\sigma$(B) &	$\mathrm{z}_\mathrm{offset}$ & incl\\
\colhead{} & \colhead{} & (\Msun) &	(K) & (K) & (MG) & (MG) & ($\mathrm{R}_\mathrm{*}$) & (°)}
\startdata
J113212.99-003036.8	&	0282-51658-0278	&	---	&	---	&	20000	&	3.19	&	0.46	&	0.04	&	21.79	\\
J114720.40-002405.7	&	0283-51584-0120	&	0.68	&	15919	&	17000	&	2.00	&	---	&	---	&	33.47	\\
J121105.25-004628.5	&	0287-52023-0253	&	1.34	&	27473	&	19500	&	2.56	&	0.01	&	-0.32	&	66.65	\\
J121635.36-002656.3	&	0288-52000-0276	&	---	&	---	&	15000	&	64.24	&	0.11	&	-0.18	&	59.27	\\
J130807.48-010117.0	&	0294-51986-0089	&	0.47	&	11757	&	18500	&	2.20	&	0.00	&	-0.21	&	72.94	\\
J144114.21+003702.3	&	0307-51663-0595	&	0.87	&	25472	&	21000	&	4.39	&	0.49	&	0.49	&	20.70	\\
J112852.88-010540.7	&	0326-52375-0565	&	---	&	---	&	18000	&	2.81	&	0.36	&	0.36	&	0.08	\\
J113431.97-031529.0	&	0327-52294-0131	&	1.24	&	14539	&	15500	&	3.00	&	0.01	&	0.34	&	16.44	\\
\enddata
\end{deluxetable}

\bibliography{sample631}{}
\bibliographystyle{aasjournal}
\end{document}